\documentclass[9pt,twocolumn,twoside]{opticajnl}
\journal{opticajournal} 

\setboolean{shortarticle}{true}

\usepackage{lineno}

\usepackage{bm}

\title{An Eigenmode Expansion Framework for Bi-Directional Multimode Nonlinear Mixing Processes}

\author{Salvador Poveda-Hospital}
\author{Vincent Girouard}
\author{Yves-Alain Peter}

\affil{Department of Engineering Physics, Polytechnique Montr\'eal, Montr\'eal, QC, H3T 1J4, Canada}

\affil[*]{salvador.poveda-hospital@polymtl.ca}

\begin{abstract}
An approach to simulating complex photonic systems is the eigenmode expansion method, 
in which scattering matrices are used to model and track the electric field in layered media. In this work, we formulate parametric nonlinear processes within the eigenmode expansion framework, enabling the simulation of nonlinear generation in varying photonic systems. This new simulation method includes both classical nonlinear processes and their quantum counterparts, including spontaneous parametric down-conversion and spontaneous four-wave mixing.
\end{abstract}

\setboolean{displaycopyright}{false} 

\begin{document}

\maketitle

\section{Introduction}
The eigenmode expansion method is based on subdividing the system into multiple cells \cite{gallagher2003eigenmode}.
Within each cell, the waveguide is approximated as invariant along the propagation direction. Maxwell's equations can thus be solved in terms of local propagation modes, characterized by a transverse field distribution and an exponential longitudinal dependence ($e^{ikz}$) \cite{Snyder1978, bures2009guided}.
Accordingly, at each longitudinal position, a set of transverse eigenmodes can be defined. The spatial part of the complex electric and magnetic fields of the $m$-th mode, at the $c$-th cell, is written as
\begin{subequations}
\begin{align}
    &\mathbf{E}^{(m,c)}(\mathbf{r})
= A^{(m,c)} \, \mathbf{e}^{(m,c)}(\mathbf{r}_\perp) \, e^{ik^{(m,c)} (z-z_c)},\\
&\mathbf{H}^{(m,c)}(\mathbf{r})
= A^{(m,c)} \, \mathbf{h}^{(m,c)}(\mathbf{r}_\perp) \, e^{ik^{(m,c)} (z-z_c)},
\end{align}
\end{subequations}
where $z_c$ denotes the longitudinal position of the entrance of the $c$-th cell, $A^{(m,c)}$ is the slowly varying modal amplitude evaluated at $z=z_c$, $\mathbf{r}_\perp=(x,y)$ is the transverse coordinates, and $k^{(m,c)}$ is the propagation constant of the  $m$-th mode in the $c$-th cell. Finally, $\mathbf{e}^{(m,c)}$ and $\mathbf{h}^{(m,c)}$ are the transverse modal fields, normalized such that each mode carries unit optical power \cite{bures2009guided, yariv1973coupled, Suhara2010-bn}, i.e.,
\begin{align}
    N = \frac{1}{2}\mathrm{Re}\iint_{\mathbb{R}^2}\left(\mathbf{e}^{(m,c)} \times \mathbf{h}^{*(m,c)}\right)\cdot \hat{\mathbf{z}}\ dA = 1.
\end{align}
Therefore, the total average power carried by the mode is simply $P = |A^{(m,c)}|^2$. Within each cell, each local eigenmode propagates independently according to
\begin{align}
\mathbf E^{(m,c)}(\mathbf{r}_\perp, z_c+\Delta z) =
\mathbf E^{(m,c)}(\mathbf{r}_\perp, z_c)e^{ik^{(m,c)}\Delta z}.
\end{align}
At the interfaces between adjacent cells, modal coupling is described through overlap integrals between eigenmodes of neighboring sections \cite{mu2008simulation}, leading to a scattering-matrix formulation of the eigenmode expansion method.
Because the electromagnetic field is expanded on a reduced modal basis, EME typically requires less memory and has a lower computational cost than fully discretized methods. This makes it particularly attractive for iterative design and optimization of integrated photonic devices \cite{oktay2024computationally, hammond2022deep}.
Its modal nature makes it particularly well suited for extensions to nonlinear optics, where interactions are governed by mode overlap and phase matching \cite{poveda2025parametric}. 

Many linear photonic devices cannot be described accurately by a purely unidirectional propagation model. Discontinuities, resonant sections, Bragg structures, mode converters, and interfaces can generate reflections and couple forward and backward propagating modes \cite{xie2020chip, memon2021recent}. As we will derive, the scattering matrix formulation is a natural framework for treating nonlinear frequency mixing in complex integrated circuits, since it can be cascaded with the linear scattering matrices of individual sections while retaining bidirectional propagation. In this work, we extend this approach by embedding the nonlinear evolution associated with difference frequency generation (DFG), sum frequency generation (SFG), and four wave mixing (FWM) into the eigenmode-expansion formalism, and by relating the same matrices to the corresponding spontaneous processes, namely spontaneous parametric down conversion (SPDC) and spontaneous four wave mixing (SFWM). This enables the calculation of both classical conversion efficiencies and quantum photon-pair generation probabilities, including their directional emission channels.

Several matrix based approaches have been proposed to model nonlinear optical propagation in layered structures \cite{albuquerque2011transfer, li2015application, huang2018new}. Bethune extended the conventional transfer matrix method to include nonlinear polarization sources, with the net generated field expressed as a sum of the contributions produced independently by the individual nonlinear layers \cite{bethune1989optical}. While this approach provides an efficient description of weak nonlinear generation, the generated fields do not mutually interact within each layer. Consequently, it does not capture the coupled signal--idler evolution responsible for parametric amplification, whose analytical solutions can exhibit hyperbolic $\sinh$ and $\cosh$ dependence on the propagation distance. More recently, the transfer matrix method was extended by constructing a nonlinear propagation matrix directly from the coupled wave equations, allowing the amplitude and phase evolution, bidirectional propagation, and parametric gain to be described within each nonlinear section \cite{poveda2025parametric}. In the present work, we build on this latter formulation but adopt a scattering matrix representation and extend the nonlinear propagation formalism to simultaneously account for multiple transverse modes, allowing nonlinear generation between any combination of signal and idler modes.

This multimode extension naturally connects the nonlinear scattering matrix formalism with the eigenmode expansion method.
The eigenmode expansion method is widely used to simulate mode coupling \cite{de2024supymode} in varying photonic structures, derived from coupled mode theory \cite{huang1994coupled, soldano1995optical}.
As derived in this work, nonlinear mixing processes exhibit analogous dynamics; however, instead of coupling between different polarizations of the modes at the same frequency, the nonlinear interaction couples modes at different frequencies.

This paper is organized as follows: Section 2 presents the scattering-matrix formulation for nonlinear mixing processes within the eigenmode-expansion framework. In Subsection 2.A, we derive the multimode formulation for DFG, where the formalism is written for $N$ signal and $N$ idler modes. We then discuss how the same scattering matrix can be interpreted in the spontaneous regime to describe SPDC, where photon pairs are generated from vacuum fluctuations. In Subsection 2.B, the approach is extended to SFG, again considering a multimode basis for the signal and idler fields. In Subsection 2.C, we apply the same framework to degenerate FWM. In this last subsection, we include self-phase modulation. We then explain how the resulting scattering matrix can be used in the SFWM regime to determine the probability amplitudes of the generated photon-pair channels, including forward, backward, and counter propagating emission.

\section{Scattering formalism for nonlinear mixing processes}

\subsection{Difference Frequency Generation}
In a longitudinally invariant waveguide, the modal amplitude $A_m$ of a mode $m$ at angular frequency $\omega_m$ driven by a nonlinear polarization field $\mathbf{P}_{NL}$ evolve along the propagation direction $z$ according to \cite{stegeman1985nonlinear, Suhara2010-bn, madsen2023mid} 
\begin{align}
    \frac{\partial  A_m}{\partial  z} = \frac{i \omega_m}{4} e^{-ik_m z}\iint_{\mathbb{R}^2}\mathbf{e}^*_m(\mathbf{r}_\perp) \cdot \mathbf{P}_{NL} (\mathbf{r})\ dA. \label{eq:chief}
\end{align}

In DFG, a pump wave at frequency $\omega_p$ and an idler wave at frequency $\omega_i$ generate a signal wave at frequency
\begin{align}
    \omega_s = \omega_p - \omega_i ~.
\end{align}
The nonlinear polarization field in \eqref{eq:chief} can be written as 
\begin{subequations}
\begin{align}
    &\mathbf{P}_{NL}(\mathbf{r};\omega_s) = \epsilon_0\chi^{(2)}:\mathbf{E}_p(\mathbf{r}) \mathbf{E}_i^*(\mathbf{r}),\\
    &\mathbf{P}_{NL}(\mathbf{r};\omega_i) = \epsilon_0\chi^{(2)}:\mathbf{E}_p(\mathbf{r}) \mathbf{E}_s^*(\mathbf{r}),\\
    &\mathbf{P}_{NL}(\mathbf{r};\omega_p) = \epsilon_0\chi^{(2)}:\mathbf{E}_s(\mathbf{r}) \mathbf{E}_i(\mathbf{r}).
\end{align}
\end{subequations}
Under the undepleated pump approximation, this yields the following set of coupled differential equations for the slowly varying amplitude of the signal, idler and pump mode:
\begin{subequations}
    \begin{align}
        &\frac{\partial A_s}{\partial z}  = \frac{i\omega_s}{c} f_{psi} e^{i\Delta k z} A_p A_i^*,\\
        &\frac{\partial A_i}{\partial z} = \frac{i\omega_i}{c} f_{psi} e^{i\Delta k z} A_p A_s^*,\\
        &\frac{\partial A_p}{\partial z} = \frac{i\omega_p}{c} f_{psi}^* e^{-i\Delta k z} A_s A_i \approx 0.
    \end{align} \label{eq:coupled_amplitude_eq}
\end{subequations}
\eqref{eq:coupled_amplitude_eq} assumes a lossless medium and Kleinman permutation symmetry of the second-order nonlinear tensor. In \eqref{eq:coupled_amplitude_eq}, $\Delta k = k_p - k_s - k_i$ is the phase mismatch and $f_{psi}$ a unitless nonlinear coupling constant, defined as 
\begin{align}
    f_{psi} = \frac{c\epsilon_0}{4}\iint_{\mathbb{R}^2} dxdy\ \chi^{(2)}_{jkl}\ \left[\mathbf{e}_p \right]_j \left[\mathbf{e}_s \right]_k^* \left[\mathbf{e}_i \right]_l^*,&& j,k,l \in \{x,y,z\}, \label{eq:coupling_constant}
\end{align}
where the double indices indicate summation. \eqref{eq:coupling_constant} is a normalized nonlinear mode overlap and accounts for waveguide geometry as, in general, $\chi^{(2)}$ is a transverse spatial profile representing the distribution of nonlinear materials in the transverse x–y plane.

The notation used in \eqref{eq:coupled_amplitude_eq} is limited when considering spatially varying structures where reflections and resonant effects may occur, since the conventional nonlinear coupled-wave equations describe only the slowly varying field amplitudes rather than the full electric field amplitudes. It is therefore convenient to work directly with electric fields to obtain equations that are compatible with the EME method. For that, we use the fact that the z-derivative of $E_m = A_me^{ik_mz}$ are related through
\begin{align}
    \frac{\partial  E_m}{\partial  z} = ik_m E_m + \frac{\partial  A_m}{\partial  z}e^{ik_mz}  ~.
\end{align}
Then, for a single signal, idler and pump mode, the coupled differential equations
written directly in terms of the electric fields are
\begin{subequations}
\begin{align}
&\frac{\partial  E_s}{\partial z} =
ik_s E_s + i\,\frac{\omega_s}{c} f_{psi} E_p E_i^* ~, \\
&\frac{\partial E_i}{\partial z} =
ik_i E_i +i\,\frac{\omega_i}{c} f_{psi} E_p E_s^* ~, \label{eq:Ep_diff_eq}\\
&\frac{\partial E_p}{\partial z} \approx ik_p E_p ~.
\end{align} \label{eq:couples_e_equations}
\end{subequations}
The first terms describe the linear propagation of the modes,
whereas the second terms describe their nonlinear coupling. The exact solution for single idler and signal modes can be found in Appendix A. \eqref{eq:Ep_diff_eq} can be solved directly for a cell starting at $z=z_c$, resulting in 
\begin{align}
    E_p(z_c + \Delta z) = E_p(z_c) e^{ik_p \Delta z}, \label{eq:solution_ep}
\end{align}
where $\Delta z = z - z_c$

We now extend the formulation of \eqref{eq:couples_e_equations} to $N$ transverse signal modes and $N$
transverse idler modes. The electric-field vectors are defined as
\begin{subequations} \label{eq:e_fields_N}
\begin{align} 
\mathbf E_s &=
\begin{pmatrix}
E_{s}^{(1)} & \cdots & E_{s}^{(N)}
\end{pmatrix}^{T}, \\
\mathbf E_i^* &=
\begin{pmatrix}
E_{i}^{(1)*} & \cdots & E_{i}^{(N)*}
\end{pmatrix}^{T}.
\end{align}
\end{subequations}

The corresponding propagation-constant matrices are
\begin{subequations} \label{eq:prop_mat}
\begin{align}
\mathbf K_s &=
\operatorname{diag}
\left(
k_{s}^{(1)},\ldots,k_{s}^{(N)}
\right), \\
\mathbf K_i &=
\operatorname{diag}
\left(
k_{i}^{(1)},\ldots,k_{i}^{(N)}
\right).
\end{align}
\end{subequations}

\begin{figure}[h]
    \centering
    \includegraphics[width=0.89\linewidth]{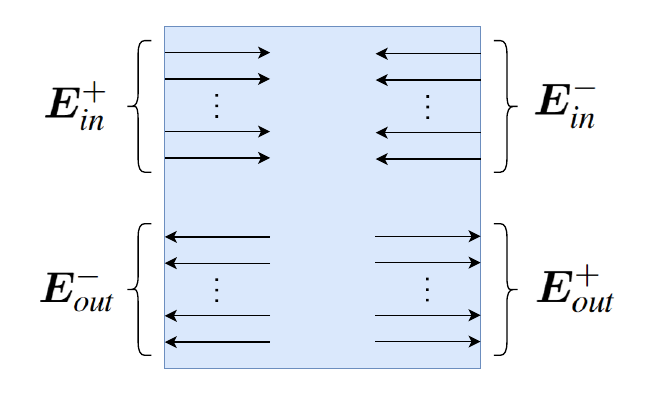}
    \caption{Schematic representation of the scattering matrix formalism
    for $N$ modes, in linear optics.}
    \label{fig:Scaterring_M}
\end{figure}

\begin{figure}[h]
    \centering
    \includegraphics[width=0.95\linewidth]{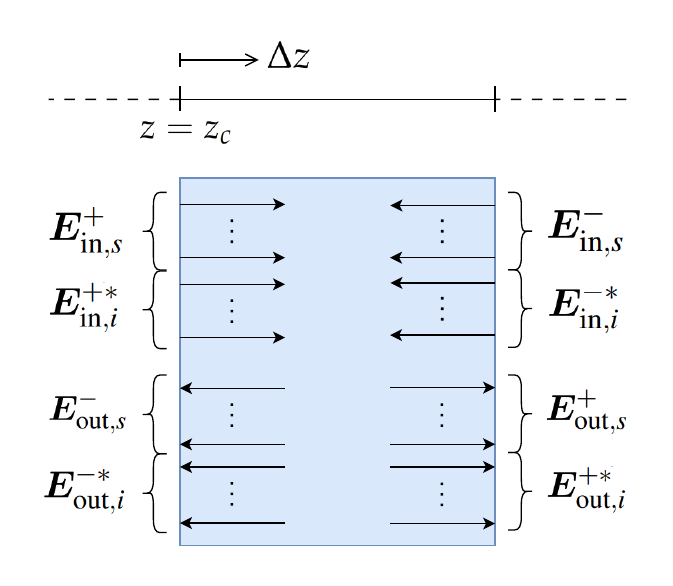}
    \caption{Schematic representation of the scattering matrix formalism
    for $N$ signal and idler modes, for linear propagation and nonlinear generation.}
    \label{fig:Scaterring_NL_M}
\end{figure}

Fig.~\ref{fig:Scaterring_M} illustrates the conventional scattering-matrix formalism for linear propagation, where the scattering matrix has dimensions $2N \times 2N$ and accounts for reflection and transmission between the supported modes.
Fig.~\ref{fig:Scaterring_NL_M} extends this formalism to include nonlinear interactions and the resulting signal and idler generation, leading to a $4N \times 4N$ scattering matrix.
These schematics provide a visual representation of the scattering-matrix framework derived in the following section. 

For a fixed pump mode $l$, the coupling coefficients defined in \eqref{eq:coupling_constant} are collected into the
$N\times N$ matrix
\begin{equation}
\label{eq:Fpsi}
\mathbf F_{psi}^{(l)}
=
\begin{pmatrix}
f_{psi}^{(l,1,1)} & \cdots & f_{psi}^{(l,1,N)} \\
\vdots & \ddots & \vdots \\
f_{psi}^{(l,N,1)} & \cdots & f_{psi}^{(l,N,N)}
\end{pmatrix},
\end{equation}
where the upper indices correspond to the mode number associated to the pump, signal and idler frequencies respectively.
Using \eqref{eq:couples_e_equations} and \eqref{eq:solution_ep}, we now define the nonlinear coupling matrices associated with a pump mode $l$ at the entrance of an EME cell located at $z_c$ as
\begin{subequations}
\begin{align}
\bm{\kappa}_s
&=
\frac{i\omega_s}{c}
E_p^{(l)}(z_c)
\mathbf F_{psi}^{(l)} ~,
\\
\bm{\kappa}_i^*
&=
-\frac{i\omega_i}{c}
E_p^{(l)*}(z_c)
\left(\mathbf F_{psi}^{(l)}\right)^{*T} ~.
\end{align}
\end{subequations}

As the pump propagates in the cell from $z_c$ to a position $z = z_c + \Delta z$, it will accumulate a linear phase $e^{i\Delta z}$, as shown in \eqref{eq:solution_ep}.
The coupled electric-field equations in the cell can thus be written in matrix form as
\begin{align}
\frac{\partial }{\partial \Delta z}
\begin{pmatrix}
\mathbf E_s(z_c+\Delta z) \\
\mathbf E_i^*(z_c+\Delta z)
\end{pmatrix}
=
\begin{pmatrix}
i\mathbf K_s
&
\bm{\kappa}_s e^{ik_p^{(l)}\Delta z}
\\
\bm{\kappa}_i^* e^{-ik_p^{(l)}\Delta z}
&
-i\mathbf K_i
\end{pmatrix}
\begin{pmatrix}
\mathbf E_s(z_c+\Delta z) \\
\mathbf E_i^*(z_c+\Delta z)
\end{pmatrix}.
\label{eq:DFG_Efield_coupled}
\end{align}

The explicit $\Delta z$ dependence of \eqref{eq:DFG_Efield_coupled} can be removed by introducing
the following rotating frame
\begin{subequations}
\begin{align}
\widetilde{\mathbf E}_s(z_c + \Delta z)
&=
e^{-ik_p^{(l)}\Delta z/2}\mathbf E_s(z_c + \Delta z) ~,
\\
\widetilde{\mathbf E}_i^*(z_c + \Delta z)
&=
e^{ik_p^{(l)}\Delta z/2}\mathbf E_i^*(z_c + \Delta z) ~.
\end{align} \label{eq:rotating_frame}
\end{subequations}
The transformation only changes their phase reference.
Differentiating the first transformed field gives
\begin{align}
\frac{\partial \widetilde{\mathbf E}_s}{\partial \Delta z}
&=
-\frac{ik_p^{(l)}}{2}
\widetilde{\mathbf E}_s
+
e^{-ik_p^{(l)}\Delta z/2}
\frac{\partial \mathbf E_s}{\partial \Delta z}
\nonumber\\
&=
i\left(
\mathbf K_s-\frac{k_p^{(l)}}{2}\mathbf I
\right)
\widetilde{\mathbf E}_s
+
\bm{\kappa}_s
\widetilde{\mathbf E}_i^* ~.
\end{align}

Similarly,
\begin{align}
\frac{\partial \widetilde{\mathbf E}_i^*}{\partial \Delta z}
&=
\frac{ik_p^{(l)}}{2}
\widetilde{\mathbf E}_i^*
+
e^{ik_p^{(l)}\Delta z/2}
\frac{\partial \mathbf E_i^*}{\partial \Delta z}
\nonumber\\
&=
\bm{\kappa}_i^*
\widetilde{\mathbf E}_s
+
i\left(
\frac{k_p^{(l)}}{2}\mathbf I-\mathbf K_i
\right)
\widetilde{\mathbf E}_i^* ~.
\end{align}

We therefore introduce the detuning matrices
\begin{subequations}
\begin{align}
\bm{\Delta}_s
&=
\frac{k_p^{(l)}}{2}\mathbf I-\mathbf K_s
\nonumber\\
&=
\operatorname{diag}
\left(
\frac{k_p^{(l)}}{2}-k_{s}^{(1)},
\ldots,
\frac{k_p^{(l)}}{2}-k_{s}^{(N)}
\right),
\\
\bm{\Delta}_i
&=
\frac{k_p^{(l)}}{2}\mathbf I-\mathbf K_i
\nonumber\\
&=
\operatorname{diag}
\left(
\frac{k_p^{(l)}}{2}-k_{i}^{(1)},
\ldots,
\frac{k_p^{(l)}}{2}-k_{i}^{(N)}
\right).
\end{align}
\end{subequations}

The coupled equations then take the $\Delta z$-independent form
\begin{align}
\frac{\partial }{\partial \Delta z}
\begin{pmatrix}
\widetilde{\mathbf E}_s \\
\widetilde{\mathbf E}_i^*
\end{pmatrix}
=
\mathbf M_{\mathrm{DFG}}
\begin{pmatrix}
\widetilde{\mathbf E}_s \\
\widetilde{\mathbf E}_i^*
\end{pmatrix},
\label{eq:MDFG_evolution}
\end{align}
where
\begin{align}
\mathbf M_{\mathrm{DFG}}
=
\begin{pmatrix}
-i\bm{\Delta}_s
&
\bm{\kappa}_s
\\
\bm{\kappa}_i^*
&
i\bm{\Delta}_i
\end{pmatrix}.
\label{eq:MDFG}
\end{align}

For a cell whose entrance is located at $z_c$, $\mathbf M_{\mathrm{DFG}}$ is constant with respect to the local propagation coordinate $\Delta z$. The solution of \eqref{eq:MDFG_evolution} over a propagation distance $\Delta z$ is therefore
\begin{align}
\begin{pmatrix}
\widetilde{\mathbf E}_s(z_c +\Delta z) \\
\widetilde{\mathbf E}_i^*(z_c +\Delta z)
\end{pmatrix}
= e^{\mathbf M_{\mathrm{DFG}}\Delta z}
\begin{pmatrix}
\widetilde{\mathbf E}_s(z_c) \\
\widetilde{\mathbf E}_i^*(z_c)
\end{pmatrix}.
\end{align}

Returning to the original electric fields using \eqref{eq:rotating_frame} gives
\begin{align}
\begin{pmatrix}
\mathbf E_s(z_c+\Delta z) \\
\mathbf E_i^*(z_c+\Delta z)
\end{pmatrix}
=
\underbrace{
\begin{pmatrix}
e^{ik_p^{(l)}\Delta z/2}\mathbf I & \bm 0 \\
\bm 0 & e^{-ik_p^{(l)}\Delta z/2}\mathbf I
\end{pmatrix}
e^{\mathbf M_{\mathrm{DFG}}\Delta z}
}_{\displaystyle \mathbf P_{\mathrm{DFG}}^{(+)}(\Delta z)}
\begin{pmatrix}
\mathbf E_s(z_c) \\
\mathbf E_i^*(z_c)
\end{pmatrix}.
\label{eq:PplusDFG}
\end{align}

Hence, the nonlinear propagation matrix for fields propagating in the
positive $z$ direction over a distance $\Delta z$ is
\begin{align}
\mathbf P_{\mathrm{DFG}}^{(+)}(\Delta z)
=
\begin{pmatrix}
e^{ik_p^{(l)}\Delta z/2}\mathbf I & \bm 0 \\
\bm 0 & e^{-ik_p^{(l)}\Delta z/2}\mathbf I
\end{pmatrix}
e^{\mathbf M_{\mathrm{DFG}}\Delta z}.
\end{align}

As a consistency check, in the absence of nonlinear coupling,
$\bm{\kappa}_s=\bm{\kappa}_i^*=\bm 0$, one obtains
\begin{align}
\mathbf P_{\mathrm{DFG}}^{(+)}(\Delta z)
\rightarrow
\begin{pmatrix}
e^{i\mathbf K_s \Delta z} & \bm 0 \\
\bm 0 & e^{-i\mathbf K_i \Delta z}
\end{pmatrix},
\end{align}
which corresponds to the usual linear propagation
$\mathbf E_s(z_c+\Delta z)=e^{i\mathbf K_s \Delta z}\mathbf E_s(z_c)$ and
$\mathbf E_i^*(z_c+\Delta z)=e^{-i\mathbf K_i \Delta z}\mathbf E_i^*(z_c)$.

The equivalent propagation matrix for fields traveling in the negative
$z$ direction, $\mathbf P_{\mathrm{DFG}}^{(-)}$, is obtained in the same
manner using the propagation constants and pump phase associated with the
negative-$z$ modes. The nonlinear propagation scattering matrix is finally written as
\begin{align}
\mathbf P_{\mathrm{DFG}}=
    \begin{pmatrix}
    \bm 0 & \mathbf P_{\mathrm{DFG}}^{(-)} \\
    \mathbf P_{\mathrm{DFG}}^{(+)}& \bm 0
    \end{pmatrix}.
\label{eq:P_Pm_Pp}
\end{align}
If we have a complex photonic system, the system can be subdivided into $C$ cells and the final solution is
\begin{equation}
\begin{pmatrix}
\bm{E}_{s,out}^- \\
\bm{E}_{i,out}^{-*} \\
\bm{E}_{s,out}^+ \\
\bm{E}_{i,out}^{+*} \\
\end{pmatrix}
=
\prod_{c=1}^C
\mathbf{S}^c \star\mathbf{P}^c
\begin{pmatrix}
\bm{E}_{s,in}^+ \\
\bm{E}_{i,in}^{+*} \\
\bm{E}_{s,in}^- \\
\bm{E}_{i,in}^{-*}
\end{pmatrix},
\label{eq:scattering_matrix}
\end{equation}
where $\star$  represents the Redheffer star product \cite{rumpf2011improved}, $\prod$ is used to denote the ordered Redheffer star product, and $\mathbf{S}$ represents the scattering matrix built from the boundary conditions between two adjacent cells \cite{eleftheriades1994some, saleh2019fundamentals}.
The associativity of the Redheffer star product is a key property in the present formulation, as it allows the global scattering matrix to be obtained through the successive combination of the scattering matrices associated with individual cells \cite{mistiri1986star}, thereby facilitating the analysis of large structures.

Spontaneous parametric down-conversion (SPDC) photon-pair sources can be characterized through the corresponding stimulated emission process, namely DFG \cite{liscidini2013stimulated, helt2012does, kulkarni2022classical}, since both processes are governed by the same nonlinear interaction and can therefore be described by the same scattering matrix.
Once the scattering matrix of the SPDC process has been determined, it provides a complete linear mapping between the incoming and outgoing field operators, thereby fully characterizing the nonlinear interaction. In the SPDC regime, where photon pairs are generated from vacuum fluctuations through the annihilation of pump photons the scattering matrix elements directly determine the probability amplitudes associated with the different pair-generation channels  \cite{Sorensen25, tassis2026parametric}. From these amplitudes, one can calculate the probabilities of generating both photons in the forward-propagating modes, both photons in the backward-propagating modes, or one photon in each propagation direction. Consequently, the scattering-matrix formalism provides a unified framework for predicting photon-pair generation rates and directionality in both classical and quantum nonlinear photonic circuits.

\subsection{Sum-frequency generation}

The eigenmode-expansion formalism can be extended directly to sum-frequency generation (SFG), where a pump wave at frequency $\omega_p$ and a signal wave at frequency $\omega_s$ generate an idler wave at frequency
\begin{equation}
\omega_i=\omega_p+\omega_s ~.
\end{equation}
The corresponding phase mismatch is
\begin{equation}
\Delta k = k_p+k_s-k_i ~.
\end{equation}
As in the DFG derivation, it is convenient to work directly with the electric fields rather than with the slowly varying amplitudes, so that we obtain electric fields compatible with the EME method. 
Under the undepleted pump approximation, the single mode electric field coupled equations are
\begin{subequations}
\begin{align}
&\frac{\partial  E_s}{\partial z} =
ik_s E_s + i\,\frac{\omega_s}{c} f_{psi}^* E_p^* E_i ~, \\
&\frac{\partial E_i}{\partial z} =
ik_i E_i +i\,\frac{\omega_i}{c} f_{psi} E_p E_s ~, \label{eq:Ep_diff_eq}\\
&\frac{\partial E_p}{\partial z} \approx ik_p E_p ~.
\end{align} \label{eq:couples_e_equations}
\end{subequations}
The first terms account for linear propagation, while the second terms describe the nonlinear exchange between signal and idler mediated by the pump.
The exact single mode solution is given in Appendix~B.
We now extend the formulation to $N$ transverse signal modes and $N$ transverse idler modes. 

Using the same definition of the transverse overlap matrix $\mathbf F_{psi}^{(l)}$ introduced for DFG, with the appropriate complex conjugation of the modal overlap for the SFG interaction, the nonlinear coupling matrices associated with pump mode $l$ at the entrance of a cell located at $z_c$ are
\begin{subequations}
\begin{align}
\bm{\kappa}_s
&= \frac{i\omega_s}{c}
E_p^{(l)*}(z_c) \left(\mathbf F_{psi}^{(l)}\right)^* ~,
\\
\bm{\kappa}_i
&= \frac{i\omega_i}{c}
E_p^{(l)}(z_c) \left(\mathbf F_{psi}^{(l)}\right)^{T} ~.
\end{align}
\end{subequations}

Using the following the rotating frame
\begin{subequations}
\begin{align}
\widetilde{\mathbf E}_s(z_c+\Delta z) &=
e^{ik_p^{(l)}\Delta z/2}\mathbf E_s(z_c+\Delta z) ~,
\\
\widetilde{\mathbf E}_i(z_c+\Delta z) &=
e^{-ik_p^{(l)}\Delta z/2}\mathbf E_i(z_c+\Delta z) ~,
\end{align}
\end{subequations}
and defining
\begin{subequations}
\begin{align}
\bm\Delta_s &=
\operatorname{diag}
\left( \frac{k_p^{(l)}}{2}+k_{s,1},\ldots, \frac{k_p^{(l)}}{2}+k_{s,N} \right), \\
\bm\Delta_i &=
\operatorname{diag}
\left( \frac{k_p^{(l)}}{2}-k_{i,1},\ldots, \frac{k_p^{(l)}}{2}-k_{i,N} \right).
\end{align}
\end{subequations}
The coupled system becomes $\Delta z$ independent,
\begin{align}
\frac{\partial }{\partial \Delta z}
\begin{pmatrix}
\widetilde{\mathbf E}_s\\
\widetilde{\mathbf E}_i
\end{pmatrix}
=
\mathbf M_{\mathrm{SFG}}
\begin{pmatrix}
\widetilde{\mathbf E}_s\\
\widetilde{\mathbf E}_i
\end{pmatrix},
\end{align}
with
\begin{align}
\mathbf M_{\mathrm{SFG}}
=
\begin{pmatrix}
i\bm\Delta_s & \bm\kappa_s\\
\bm\kappa_i & -i\bm\Delta_i
\end{pmatrix}.
\end{align}
For a cell whose entrance is located at $z_c$, $\mathbf M_{\mathrm{SFG}}$ is constant with respect to $\Delta z$, and therefore
\begin{align}
\begin{pmatrix}
\widetilde{\mathbf E}_s(\Delta z)\\
\widetilde{\mathbf E}_i(\Delta z)
\end{pmatrix}
=
e^{\mathbf M_{\mathrm{SFG}}\Delta z}
\begin{pmatrix}
\widetilde{\mathbf E}_s(0)\\
\widetilde{\mathbf E}_i(0)
\end{pmatrix}.
\end{align}
Returning to the original electric fields gives
\begin{align}
\begin{pmatrix}
\mathbf E_s(z_c+\Delta z)\\
\mathbf E_i(z_c+\Delta z)
\end{pmatrix}
=
\underbrace{
\begin{pmatrix}
e^{-ik_p^{(l)}\Delta z/2}\mathbf I & \bm 0\\
\bm 0 & e^{ik_p^{(l)}\Delta z/2}\mathbf I
\end{pmatrix}
e^{\mathbf M_{\mathrm{SFG}}\Delta z}
}_{\displaystyle \mathbf P_{\mathrm{SFG}}^{(+)}(\Delta z)}
\begin{pmatrix}
\mathbf E_s(z)\\
\mathbf E_i(z)
\end{pmatrix}.
\label{eq:PplusSFG}
\end{align}
Thus,
\begin{align}
\mathbf P_{\mathrm{SFG}}^{(+)}(\Delta z)
=
\begin{pmatrix}
e^{-ik_p^{(l)}\Delta z/2}\mathbf I & \bm 0\\
\bm 0 & e^{ik_p^{(l)}\Delta z/2}\mathbf I
\end{pmatrix}
e^{\mathbf M_{\mathrm{SFG}}\Delta z} ~.
\end{align}
The negative-$z$ matrix $\mathbf P_{\mathrm{SFG}}^{(-)}$ is obtained analogously, and the complete $4N\times4N$ nonlinear propagation scattering matrix is constructed in the same way as for DFG in ~\eqref{eq:P_Pm_Pp}.

\subsection{Four-wave mixing}

The same scattering matrix formulation can be applied to four-wave mixing (FWM). We further extend the derivation to also include the effects of self-phase modulation (SPM) and cross-phase modulation (XPM) in the undepleted pump regime. In FWM, two pump modes labeled $p$ and $p'$ at frequencies $\omega_p$ and $\omega_{p'}$ interact with an idler field at frequency $\omega_i$ to generate a signal according to
\begin{equation}
\omega_s=\omega_p+\omega_{p'}-\omega_i ~,
\end{equation}
with phase mismatch
\begin{equation}
\Delta k=k_p+k_{p'}-k_s-k_i ~.
\end{equation}
Under the undepleted pump approximation, the nonlinear polarization field $\mathbf{P}_{NL}$ at the relevant frequencies can be approximated by
\begin{subequations}
    \begin{align}
\mathbf{P}_{NL}(\mathbf{r}, \omega_s) &\approx \frac{3}{2}\epsilon_0\chi^{(3)}:\left(\mathbf{E}_p \mathbf{E}_p^*\mathbf{E}_s + \mathbf{E}_{p'} \mathbf{E}_{p'}^*\mathbf{E}_s + \mathbf{E}_p \mathbf{E}_{p'}\mathbf{E}_i^*\right),\label{eq:P_s_FWM}\\
\mathbf{P}_{NL}(\mathbf{r}, \omega_i) &\approx \frac{3}{2}\epsilon_0\chi^{(3)}:\left(\mathbf{E}_p \mathbf{E}_p^*\mathbf{E}_i + \mathbf{E}_{p'} \mathbf{E}_{p'}^*\mathbf{E}_i + \mathbf{E}_p \mathbf{E}_{p'}\mathbf{E}_s^*\right),\label{eq:P_i_FWM}\\
\mathbf{P}_{NL}(\mathbf{r}, \omega_p) &\approx \frac{3}{4}\epsilon_0\chi^{(3)}:\left(\mathbf{E}_p \mathbf{E}_p^*\mathbf{E}_p + 2\mathbf{E}_{p'} \mathbf{E}_{p'}^*\mathbf{E}_p\right),\label{eq:P_p_FWM}\\
\mathbf{P}_{NL}(\mathbf{r}, \omega_{p'}) &\approx \frac{3}{4}\epsilon_0\chi^{(3)}:\left(\mathbf{E}_{p'} \mathbf{E}_{p'}^*\mathbf{E}_{p'} + 2\mathbf{E}_p \mathbf{E}_p^*\mathbf{E}_i\right), \label{eq:P_p'_FWM}
    \end{align}
\end{subequations}
where we only kept the SPM and XPM interactions that arise from the two strong pump fields \cite{nonlinear-fiber-optics}. The coefficients come from the different combinations of the total electric field frequency components.
Using the expressions for the nonlinear polarizations at $\omega_p$ (\eqref{eq:P_p_FWM}) and at $\omega_{p'}$ (\eqref{eq:P_p'_FWM}) into \eqref{eq:chief} yields two uncoupled equations for the pump modal amplitudes:
\begin{subequations}
    \begin{align}
        &\frac{\partial A_p}{\partial z} = i\underbrace{\frac{\omega_p}{c}\left(f_{pp}|A_p|^2 + 2f_{pp'}|A_{p'}|^2\right)}_{\gamma_p} A_p = i\gamma_pA_p~,\label{eq:diff_eq_Ap_FWM}\\
        &\frac{\partial A_{p'}}{\partial z} = i\underbrace{\frac{\omega_{p'}}{c}\left(f_{p'p'}|A_{p'}|^2 + 2f_{p'p}|A_p|^2\right)}_{\gamma_{p'}} A_{p'} = i\gamma_{p'}A_{p'}\label{eq:diff_eq_Ap'_FWM}~,
    \end{align}
\end{subequations}
where $|A_p|^2$ and $|A_{p'}|^2$ are the total average power carried by the pump modes and are constant within a cell with respect to $z$ under the undepleted pump approximation. The nonlinear coupling constants $f_{\mu \nu}$ are defined as 
\begin{align}
    f_{\mu \nu} = \frac{3}{16}c\epsilon_0 \iint_{\mathbb{R}^2} dxdy\ \chi^{(3)}_{jklm} [\mathbf{e}_\mu]^*_j[\mathbf{e}_\mu]_k[\mathbf{e}_\nu]^*_l[\mathbf{e}_\nu]_m~,\\
    \nonumber j,k,l,m \in \{x,y,z\}~.
\end{align}
where the double indices indicate summation.

\eqref{eq:diff_eq_Ap_FWM} and \eqref{eq:diff_eq_Ap'_FWM} can be solved directly, and the solutions are expressed in terms of the full electric field by including the longitudinal phase factor $e^{ikz}$. For a cell starting at $z_c$, the electric field of the pumps over a propagation distance $\Delta z$ is therefore
\begin{subequations}
    \begin{align}
        &E_p(z_c + \Delta z) = E_p(z_c)e^{i(k_p + \gamma_p)\Delta z},\\
        &E_{p'}(z_c + \Delta z) = E_{p'}(z_c)e^{i(k_{p'} + \gamma_{p'})\Delta z}.
    \end{align} \label{eq:solution_ep_FWM}
\end{subequations} 
The effects of SPM and XPM on the evolution of the pumps are entirely included into the constants $\gamma_p$ and $\gamma_{p'}$.

Using \eqref{eq:P_s_FWM} and \eqref{eq:P_i_FWM} into \eqref{eq:chief}, the coupled-amplitude equations for given signal and idler modes can be written as
\begin{subequations}
    \begin{align}
        &\frac{\partial A_s}{\partial z} = i\gamma_sA_s + i\frac{2\omega_s}{c}f_{sipp'}A_pA_{p'}A_i^* e^{i\Delta k z},\\
        &\frac{\partial A_i}{\partial z} = i\gamma_iA_i + i\frac{2\omega_i}{c}f_{sipp'}A_pA_{p'}A_s^* e^{i\Delta k z},
    \end{align}\label{eq:coupled_amplitude_FWM}
\end{subequations}
where the constants $\gamma_s$ and $\gamma_i$ include the effects of cross-phase modulation with the pump fields and are given by
\begin{subequations}
    \begin{align}
        &\gamma_s = \frac{2\omega_s}{c}\left(f_{sp}|A_p|^2 + f_{sp'}|A_{p'}|^2\right),\\
        &\gamma_i = \frac{2\omega_i}{c}\left(f_{ip}|A_p|^2 + f_{ip'}|A_{p'}|^2\right).
    \end{align}
\end{subequations}
The nonlinear coupling constant $f_{sipp'}$ is given by 
\begin{align}
    f_{sipp'} = \frac{3}{16}c\epsilon_0\iint_{\mathbb{R}^2}dxdy\ \chi^{(3)}_{jklm} [\mathbf{e}_s]^*_j[\mathbf{e}_i]^*_k[\mathbf{e}_p]_l[\mathbf{e}_{p'}]_m~,\\
    \nonumber j,k,l,m \in \{x,y,z\}~.
\end{align}

Finally, like for DFG, the coupled amplitude equations in \eqref{eq:coupled_amplitude_FWM} are written in terms of the electric fields, which gives
\begin{subequations}
    \begin{align}
        &\frac{\partial E_s}{\partial z} = i\left(k_s + \gamma_s\right)E_s + i\frac{2\omega_s}{c}f_{sipp'}E_pE_{p'}E_i^*~,\\
        &\frac{\partial E_i}{\partial z} = i\left(k_i + \gamma_i\right)E_i + i\frac{2\omega_i}{c}f_{sipp'}E_pE_{p'}E_s^*~.
    \end{align}\label{eq:coupled_e_equations_FWM}
\end{subequations}
The product $E_pE_{p'}$ \eqref{eq:coupled_e_equations_FWM} acts as an effective pump field, and the constant $k_{s,p} + \gamma_{s,p}$ acts as an effective propagation constant, which makes the structure of the FWM equations analogous to DFG.

We now extend the formulation of the coupled equations to $N$ transverse signal and idler modes. For a fixed pair of pump modes $(l,m)$, the FWM coupling coefficients are collected into the $N\times N$ matrix
\begin{equation}
\label{eq:Fppsi}
\mathbf F_{sipp'}^{(l,m)}
=
\begin{pmatrix}
f_{sipp'}^{(1,1,l,m)} & \cdots & f_{sipp'}^{(1,N,l,m)}\\
\vdots & \ddots & \vdots\\
f_{sipp'}^{(N,1,l,m)} & \cdots & f_{sipp'}^{(N,N,l,m)}
\end{pmatrix}~,
\end{equation}
where the upper indices correspond to the mode number associated with the signal, idler, and pump frequencies, respectively.
In a similar way, the XPM coupling coefficients are grouped in the $N\times N$ matrices 
\begin{subequations}
    \begin{align}
        \mathbf{F}_{sp}^{(l)} &= \mathrm{diag}\left[f_{sp}^{(1,l)}, \dots, f_{sp}^{(N,l)}\right]~,\\
        \mathbf{F}_{sp'}^{(m)} &= \mathrm{diag}\left[f_{sp'}^{(1,m)}, \dots, f_{sp'}^{(N,m)}\right]~,\\
        \mathbf{F}_{ip}^{(l)} &= \mathrm{diag}\left[f_{ip}^{(1,l)}, \dots, f_{ip}^{(N,l)}\right]~,\\
        \mathbf{F}_{ip'}^{(m)} &= \mathrm{diag}\left[f_{ip'}^{(1,m)}, \dots, f_{ip'}^{(N,m)}\right]~.
    \end{align}
\end{subequations}

We now define the XPM coupling matrices, evaluated at the entrance of a cell located at $z_c$, as
\begin{subequations}
    \begin{align}
        \bm\Gamma_s^{(l,m)} = \frac{2\omega_s}{c}\left(\mathbf{F}_{sp}^{(l)}|E_p(z_c)|^2 + \mathbf{F}_{sp'}^{(m)}|E_{p'}(z_c)|^2\right),\\
        \bm\Gamma_i^{(l,m)} = \frac{2\omega_i}{c}\left(\mathbf{F}_{ip}^{(l)}|E_p(z_c)|^2 + \mathbf{F}_{ip'}^{(m)}|E_{p'}(z_c)|^2\right).
    \end{align}
\end{subequations}
From \eqref{eq:solution_ep_FWM} and \eqref{eq:coupled_e_equations_FWM}, we also define the nonlinear coupling matrices for this pump-mode pair, evaluated at the entrance of a cell located at $z_c$ as
\begin{subequations}
\begin{align}
\bm\kappa_s
&=
\frac{i2\omega_s}{c}
E_p^{(l)}(z_c)E_{p'}^{(m)}(z_c)
\mathbf F_{sipp'}^{(l,m)} ~,
\\
\bm\kappa_i^{*}
&=
-\frac{i2\omega_i}{c}
E_p^{(l)*}(z_c)E_{p'}^{(m)*}(z_c)
\left(\mathbf F_{sipp'}^{(l,m)}\right)^{*T} ~.
\end{align}
\end{subequations}

We now express the coupled electric field equations of \eqref{eq:coupled_e_equations_FWM} in matrix form. Let
\begin{align}
K_{p}^{(l,m)}=k_p^{(l)} + \gamma_p^{(l,m)}+k_{p'}^{(m)} + \gamma_{p'}^{(l,m)}
\end{align}
be the effective propagation constant of the pump.
According to \eqref{eq:solution_ep_FWM}, for a fixed pump-mode pair $(l,m)$, we have
\begin{align}
E_p^{(l)}(z_c+\Delta z)E_{p'}^{(m)}(z_c+\Delta z)
=
E_p^{(l)}(z_c)E_{p'}^{(m)}(z_c)e^{iK_{p}^{(l,m)}\Delta z} ~.
\end{align}
Therefore, the coupled electric field equations in \eqref{eq:coupled_e_equations_FWM} are expressed as 
\begin{align}
\begin{split}
    &\frac{\partial }{\partial \Delta z}
\begin{pmatrix}
\mathbf E_s(z_c+\Delta z) \\
\mathbf E_i^*(z_c+\Delta z)
\end{pmatrix}
=
\\&\ \begin{pmatrix}
i(\mathbf K_s + \bm\Gamma_s)
&
\bm{\kappa}_s e^{iK_p^{(l,m)}\Delta z}
\\
\bm{\kappa}_i^* e^{-iK_p^{(l,m)}\Delta z}
&
-i(\mathbf K_i + \bm \Gamma_i^*)
\end{pmatrix}
\begin{pmatrix}
\mathbf E_s(z_c+\Delta z) \\
\mathbf E_i^*(z_c+\Delta z)
\end{pmatrix}.
\end{split}
\label{eq:FWM_Efield_coupled}
\end{align}
The explicit $\Delta z$ dependence is removed with the following rotating frame
\begin{subequations}
\begin{align}
\widetilde{\mathbf E}_s(z_c +\Delta z)
&=
e^{-iK_p^{(l,m)}\Delta z/2}\mathbf E_s(z_c+\Delta z) ~,
\\
\widetilde{\mathbf E}_i^*(z_c +\Delta z)
&=
e^{iK_p^{(l,m)}\Delta z/2}\mathbf E_i^*(z_c+\Delta z) ~.
\end{align}
\end{subequations}
We define the corresponding detuning matrices
\begin{subequations}
\begin{align}
\bm\Delta_s^{(l,m)}
&=
\frac{K_p^{(l,m)}}{2}\mathbf I-\mathbf K_s - \bm \Gamma_s^{(l,m)}
\\
\bm\Delta_i^{(l,m)}
&=
\frac{K_p^{(l,m)}}{2}\mathbf I-\mathbf K_i - \bm \Gamma_i^{(l,m)}
\end{align}
\end{subequations}
The transformed fields thus satisfy
\begin{align}
\frac{\partial }{\partial \Delta z}
\begin{pmatrix}
\widetilde{\mathbf E}_s\\
\widetilde{\mathbf E}_i^*
\end{pmatrix}
=
\mathbf M_{\mathrm{FWM}}^{(l,m)}
\begin{pmatrix}
\widetilde{\mathbf E}_s\\
\widetilde{\mathbf E}_i^*
\end{pmatrix},
\end{align}
where
\begin{align}
\mathbf M_{\mathrm{FWM}}^{(l,m)}
=
\begin{pmatrix}
-i\bm\Delta_s^{(l,m)} & \bm\kappa_s^{(l,m)}\\
\bm\kappa_i^{(l,m)*} & i\bm\Delta_i^{(l,m)}
\end{pmatrix}.
\end{align}
Within one EME cell, the solution for this pump-mode pair is therefore
\begin{align}
\begin{pmatrix}
\widetilde{\mathbf E}_s(z_c +\Delta z)\\
\widetilde{\mathbf E}_i^*(z_c +\Delta z)
\end{pmatrix}
=
e^{\mathbf M_{\mathrm{FWM}}^{(l,m)}\Delta z}
\begin{pmatrix}
\widetilde{\mathbf E}_s(z_c)\\
\widetilde{\mathbf E}_i^*(z_c)
\end{pmatrix}.
\end{align}
Returning to the original fields gives the positive-$z$ nonlinear propagation matrix
\begin{align}
\mathbf P_{\mathrm{FWM}}^{(+),(l,m)}(\Delta z)
=
\begin{pmatrix}
e^{iK_p^{(l,m)}\Delta z/2}\mathbf I & \bm 0\\
\bm 0 & e^{-iK_p^{(l,m)}\Delta z/2}\mathbf I
\end{pmatrix}
e^{\mathbf M_{\mathrm{FWM}}^{(l,m)}\Delta z} ~.
\label{eq:PplusFWM}
\end{align}

The equivalent matrix $\mathbf P_{\mathrm{FWM}}^{(-)}$ for negative-$z$ propagation is obtained analogously. The complete $4N\times4N$ nonlinear propagation scattering matrix is then built as in ~\eqref{eq:P_Pm_Pp} and can be combined with the linear EME interface matrices through the Redheffer star product.

Once the scattering matrix of the four-wave mixing process has been determined, it provides a complete linear mapping between the incoming and outgoing field operators, thereby fully characterizing the nonlinear interaction. In the spontaneous four-wave mixing (SFWM) regime \cite{Azzini12}, where photon pairs are generated from vacuum fluctuations, the scattering matrix can be used to calculate the probabilities of the emitted photon pairs \cite{tassis2026parametric}. In particular, the matrix elements directly determine the probability amplitudes associated with the different pair-generation channels. From these amplitudes, one can calculate the probabilities of generating both photons in the forward-propagating modes, both photons in the backward-propagating modes, or one photon in each propagation direction. Consequently, the scattering-matrix formalism provides a unified framework for predicting photon-pair generation rates and directionality in both classical and quantum nonlinear photonic circuits.

\section{Conclusion}
We have developed a multimode scattering matrix formulation 
for difference frequency generation, spontaneous parametric down conversion, sum frequency generation, four wave mixing and spontaneous four wave mixing.
This scattering matrix can be imcorporated into eigenmode expansion solvers to simulate nonlinear mixing processes in arbitrary photonic circuits.
Optimization algorithms can be implemented to tailor the design to specific purposes. 
This method has the potential to accelerate research and development in classical and quantum nonlinear photonic circuits simulations.

\section*{Appendix A} \label{Annex1}
For single mode DFG, the electric field analytical solution is
\begin{subequations}
\begin{align}
E_s(z)
&=
e^{i\left(k_s+\frac{\Delta k}{2}\right)z}
\Bigg[
\left(
\cosh(g z)
-i\frac{\Delta k}{2g}\sinh(g z)
\right)E_s(0)
\nonumber\\
&\hspace{2.0cm}
+\frac{\kappa_s}{g}
\sinh(g z)\,E_i^*(0)
\Bigg],
\\
E_i^*(z)
&=
e^{-i\left(k_i+\frac{\Delta k}{2}\right)z}
\Bigg[
\frac{\kappa_i^*}{g}
\sinh(g z)\,E_s(0)
\nonumber\\
&\hspace{2.0cm}
+\left(
\cosh(g z)
+i\frac{\Delta k}{2g}\sinh(g z)
\right)E_i^*(0)
\Bigg].
\end{align}
\end{subequations}
where the nonlinear gain is
\begin{align}
g = \sqrt{ \kappa_s\kappa_i^* -\frac{\Delta k^2}{4} } .
\end{align}

\section*{Appendix B} \label{Annex2}
For single mode SFG, the electric field analytical solution is
\begin{subequations}
\begin{align}
E_s(z)
&=
e^{i\left(k_s-\frac{\Delta k}{2}\right)z}
\Bigg[
\left(
\cos(\Omega z)
+i\frac{\Delta k}{2\Omega}\sin(\Omega z)
\right)E_s(0)
\nonumber\\
&\hspace{2.0cm}
+\frac{\kappa_s}{\Omega}
\sin(\Omega z)\,E_i(0)
\Bigg],
\\
E_i(z)
&=
e^{i\left(k_i+\frac{\Delta k}{2}\right)z}
\Bigg[
\frac{\kappa_i}{\Omega}
\sin(\Omega z)\,E_s(0)
\nonumber\\
&\hspace{2.0cm}
+\left(
\cos(\Omega z)
-i\frac{\Delta k}{2\Omega}\sin(\Omega z)
\right)E_i(0)
\Bigg].
\end{align}
\end{subequations}
where the nonlinear gain is
\begin{align}
\Omega
=
\sqrt{
\frac{\Delta k^2}{4}
-\kappa_s\kappa_i
}.
\end{align}








\begin{backmatter}
\bmsection{Funding}  Y.-A. P. acknowledge support from the Ministère de l’Économie et de l’Innovation du Québec



\smallskip

\noindent Here are examples of disclosures:


\bmsection{Disclosures} The authors declare no conflicts of interest.

\end{backmatter}

\bibliography{sample}

@article{Snyder1978,
author = {Allan W. Snyder and William R. Young},
journal = {J. Opt. Soc. Am.},
number = {3}, pages = {297--309},
publisher = {Optica Publishing Group},
title = {Modes of optical waveguides},
volume = {68}, month = {Mar},
year = {1978},
doi = {10.1364/JOSA.68.000297},
}

@article{mistiri1986star,
  title={The star-product and its algebraic properties},
  author={Mistiri, F and Wang, Alan P},
  journal={Journal of the Franklin Institute},
  volume={321},  number={1},  pages={21--38},
  year={1986},
  publisher={Elsevier}
}

@article{bethune1989optical,
  title={Optical harmonic generation and mixing in multilayer media: analysis using optical transfer matrix techniques},
  author={Bethune, DS},
  journal={Journal of the Optical society of America B},
  volume={6},  number={5},  pages={910--916},
  year={1989},
  publisher={Optical Society of America}
}

@article{eleftheriades1994some,
  title={Some important properties of waveguide junction generalized scattering matrices in the context of the mode matching technique},
  author={Eleftheriades, George V and Omar, Abbas S and Katehi, Linda PB and Rebeiz, Gabriel M},
  journal={IEEE Transactions on Microwave Theory and Techniques},
  volume={42},  number={10},  pages={1896--1903},
  year={1994},
  publisher={IEEE}
}

@article{huang1994coupled,
  title={Coupled-mode theory for optical waveguides: an overview},
  author={Huang, Wei-Ping},
  journal={Journal of the Optical Society of America A},
  volume={11},  number={3},  pages={963--983},
  year={1994},
  publisher={Optical Society of America}
}

@article{soldano1995optical,
  title={Optical multi-mode interference devices based on self-imaging: principles and applications},
  author={Soldano, Lucas B and Pennings, Erik CM},
  journal={Journal of lightwave technology},
  volume={13},  number={4},  pages={615--627},
  year={1995},
  publisher={IEEE}
}

@inproceedings{gallagher2003eigenmode,
  title={Eigenmode expansion methods for simulation of optical propagation in photonics: pros and cons},
  author={Gallagher, Dominic FG and Felici, Thomas P},
  booktitle={Integrated optics: devices, materials, and technologies VII},
  volume={4987},  pages={69--82},
  year={2003},
  organization={SPIE}
}

@article{mu2008simulation,
  title={Simulation of three-dimensional waveguide discontinuities by a full-vector mode-matching method based on finite-difference schemes},
  author={Mu, Jianwei and Huang, Wei-Ping},
  journal={Optics express},
  volume={16},  number={22},  pages={18152--18163},
  year={2008},
  publisher={Optical Society of America}
}

@book{bures2009guided,
  title={Guided optics: Optical fibers and all-fiber components},
  author={Bures, Jacques},
  year={2009},
  publisher={John Wiley \& Sons}
}

@article{albuquerque2011transfer,
  title={Transfer matrix and fourier transform methods for simulation of second-order nonlinear interactions in a PPLN waveguide},
  author={Albuquerque, Andr{\'e} AC and Drummond, Miguel V and Nogueira, Rog{\'e}rio N},
  journal={Journal of lightwave technology},
  volume={29},  number={24},  pages={3764--3771},
  year={2011},
  publisher={OSA}
}

@article{rumpf2011improved,
  title={Improved formulation of scattering matrices for semi-analytical methods that is consistent with convention},
  author={Rumpf, Raymond C},
  journal={Progress In Electromagnetics Research B},
  volume={35},  pages={241--261},
  year={2011},
  publisher={EMW Publishing}
}

@article{Azzini12,
author = {Stefano Azzini and Davide Grassani and Matteo Galli and Lucio Claudio Andreani and Marc Sorel and Michael J. Strain and L. G. Helt and J. E. Sipe and Marco Liscidini and Daniele Bajoni},
journal = {Opt. Lett.},
number = {18},pages = {3807--3809},
publisher = {Optica Publishing Group},
title = {From classical four-wave mixing to parametric fluorescence in silicon microring resonators},
volume = {37},month = {Sep},year = {2012},
doi = {10.1364/OL.37.003807},
}

@article{helt2012does,
  title={How does it scale? Comparing quantum and classical nonlinear optical processes in integrated devices},
  author={Helt, Lukas G and Liscidini, Marco and Sipe, John E},
  journal={Journal of the Optical Society of America B},
  volume={29},  number={8},  pages={2199--2212},
  year={2012},
  publisher={Optical Society of America}
}

@article{liscidini2013stimulated,
  title={Stimulated emission tomography},
  author={Liscidini, Marco and Sipe, JE},
  journal={Physical review letters},
  volume={111},  number={19},  pages={193602},
  year={2013},
  publisher={APS}
}

@article{li2015application,
  title={Application of transfer matrix method to second-harmonic generation in nonlinear photonic bandgap structures: oblique incidence},
  author={Li, Han and Haus, Joseph W and Banerjee, Partha P},
  journal={Journal of the Optical Society of America B},
  volume={32},  number={7},  pages={1456--1462},  year={2015},
  publisher={Optical Society of America}
}

@article{huang2018new,
  title={New transfer-matrix method for frequency conversion in nonlinear multilayered structures based on coupled-amplitude equations},
  author={Huang, Jin jer and Feng, Qian and Zhang, Xin Lu and Zhang, Liu Yang},
  journal={Journal of the Optical Society of America B},
  volume={36},  number={1},  pages={26--34},
  year={2018},
  publisher={Optical Society of America}
}

@book{saleh2019fundamentals,
  title={Fundamentals of photonics, 2 volume set},
  author={Saleh, Bahaa EA and Teich, Malvin Carl},
  year={2019},
  publisher={john Wiley \& sons}
}

@book{nonlinear-fiber-optics,
	title = {Nonlinear Fiber Optics (6th Edition)},
	author = {Agrawal, Govind P.},
	year = {2019},
	url = {https://app.knovel.com/hotlink/toc/id:kpNFOE0012/nonlinear-fiber-optics/nonlinear-fiber-optics},
	isbn = {978-0-1281-7042-7},
	publisher = {Elsevier},
}

@article{xie2020chip,
  title={On-chip Fabry--P{\'e}rot Bragg grating cavity enhanced four-wave mixing},
  author={Xie, Shengjie and Zhang, Yang and Hu, Yiwen and Veilleux, Sylvain and Dagenais, Mario},
  journal={ACS Photonics},
  volume={7},  number={4},  pages={1009--1015},
  year={2020},
  publisher={ACS Publications}
}

@article{memon2021recent,
  title={Recent advances in mode converters for a mode division multiplex transmission system},
  author={Memon, Areez K and Chen, Kai X},
  journal={Opto-Electronics Review},
  volume={29},
  year={2021}
}

@article{kulkarni2022classical,
  title={Classical model of spontaneous parametric down-conversion},
  author={Kulkarni, Girish and Rioux, Jeremy and Braverman, Boris and Chekhova, Maria V and Boyd, Robert W},
  journal={Physical Review Research},
  volume={4},  number={3},  pages={033098},
  year={2022},
  publisher={APS}
}

@article{hammond2022deep,
  title={Deep learning-enhanced, open-source eigenmode expansion},
  author={Hammond, Ian M and Hammond, Alec M and Camacho, Ryan M},
  journal={Optics Letters},
  volume={47},  number={6},  pages={1383--1386},
  year={2022},
  publisher={Optica Publishing Group}
}

@article{oktay2024computationally,
  title={Computationally Efficient Nanophotonic Design Through Data-Driven Eigenmode Expansion},
  author={Oktay, Mehmet Can and Gorgulu, Kazim and Magden, Emir Salih},
  journal={Journal of Lightwave Technology},
  volume={42},  number={22},  pages={7894--7902},
  year={2024}
}

@article{de2024supymode,
  title={SuPyMode: an open-source library for design and optimization of fiber optic components},
  author={de Sivry-Houle, Martin Poinsinet and Becerra Deana, Rodrigo Itzamna and Virally, St{\'e}phane and Godbout, Nicolas and Boudoux, Caroline},
  journal={Optics Continuum},
  volume={3},  number={2},  pages={242--255},
  year={2024},
  publisher={Optica Publishing Group}
}

@article{poveda2025parametric,
	author={Poveda Hospital, Salvador and Tassis, Tadeu and Peter, Yves-Alain and Quesada, Nicolás and Houde, Martin},
	title={Parametric processes in nonlinear structures with reflections: a transfer matrix method approach},
	journal={Journal of Optics},
	url={http://iopscience.iop.org/article/10.1088/2040-8986/ae80e0},
	year={2026}
}

@article{Sorensen25,
author = {Nicholas J. Sorensen and Vitaliy Sultanov and Maria V. Chekhova},
journal = {Opt. Express},
number = {6},pages = {13946--13960},
publisher = {Optica Publishing Group},
title = {A simple model for entangled photon generation in resonant structures},
volume = {33}, month = {Mar},
year = {2025},
doi = {10.1364/OE.555205},
}

@article{tassis2026parametric,
  title={Parametric processes in nonlinear structures with reflections: an asymptotic-field approach},
  author={Tassis, Tadeu and Poveda-Hospital, Salvador and Quesada, Nicol{\'a}s and Houde, Martin},
  journal={Journal of Physics: Photonics},
  volume={8},  number={2},  pages={023001},
  year={2026},
  publisher={IOP Publishing}
}

@BOOK{Suhara2010-bn,
  title     = "Waveguide nonlinear-optic devices",
  author    = "Suhara, Toshiaki and Fujimura, Masatoshi",
  publisher = "Springer",
  series    = "Springer Series in Photonics",
  month     =  dec,
  year      =  2010,
  address   = "Berlin, Germany",
  language  = "en"
}

@article{yariv1973coupled,
  title={Coupled-mode theory for guided-wave optics},
  author={Yariv, Amnon},
  journal={IEEE Journal of Quantum Electronics},
  volume={9},
  number={9},
  pages={919--933},
  year={1973},
  publisher={IEEE}
}

@article{stegeman1985nonlinear,
  title={Nonlinear integrated optics},
  author={Stegeman, George I and Seaton, Colin T},
  journal={Journal of applied physics},
  volume={58},
  number={12},
  pages={R57--R78},
  year={1985},
  publisher={American Institute of Physics}
}

@article{madsen2023mid,
  title={Mid-infrared difference-frequency generation in AlGaAs-on-insulator waveguides},
  author={Madsen, Magnus L and Ulsig, Emil Z and Folsach, Sebastian and Godoy, Pedro H and Stanton, Eric J and Volet, Nicolas},
  journal={Journal of the Optical Society of America B},
  volume={40},
  number={7},
  pages={1742--1748},
  year={2023},
  publisher={Optica Publishing Group}
}



\ifthenelse{\equal{\journalref}{aop}}{%
\section*{Author Biographies}
\begingroup
\setlength\intextsep{0pt}
\begin{minipage}[t][6.3cm][t]{1.0\textwidth} 
  \begin{wrapfigure}{L}{0.25\textwidth}
    \includegraphics[width=0.25\textwidth]{john_smith.eps}
  \end{wrapfigure}
  \noindent
  {\bfseries John Smith} received his BSc (Mathematics) in 2000 from The University of Maryland. His research interests include lasers and optics.
\end{minipage}
\begin{minipage}{1.0\textwidth}
  \begin{wrapfigure}{L}{0.25\textwidth}
    \includegraphics[width=0.25\textwidth]{alice_smith.eps}
  \end{wrapfigure}
  \noindent
  {\bfseries Alice Smith} also received her BSc (Mathematics) in 2000 from The University of Maryland. Her research interests also include lasers and optics.
\end{minipage}
\endgroup
}{}

\end{document}